\pdfoutput=1
\documentclass[10pt,journal]{IEEEtran}
\usepackage{amsmath,graphicx}
\usepackage{times}
\usepackage[utf8]{inputenc}
\usepackage{cite}
\usepackage{floatrow}
\renewcommand*{\thesubsubsection}{\Alph{subsubsection}}
\usepackage{amsmath,amssymb,amsfonts,stfloats}
\usepackage{tabularx}
\usepackage{graphicx}
\usepackage{textcomp}
\usepackage{floatrow}
\usepackage{xcolor}
\usepackage[hyphens]{url}
\usepackage{caption}
\usepackage{booktabs}

\usepackage{multicol}
\usepackage{multirow}
\usepackage{acronym}
\usepackage{url}
\usepackage{mathtools}
\usepackage{algorithm}
\usepackage{algpseudocode} 
\usepackage{enumitem}
\usepackage[hang,flushmargin]{footmisc}
\usepackage[caption=false,font=footnotesize]{subfig}
\usepackage[acronyms,nonumberlist,nopostdot,nomain,nogroupskip,acronymlists={hidden}]{glossaries}
\newacronym{3gpp}{3GPP}{3rd Generation Partnership Project}
\newacronym{4g}{4G}{4th generation}
\newacronym{5g}{5G}{5th generation}
\newacronym{6g}{6G}{6th generation}
\newacronym{5gc}{5GC}{5G Core}
\newacronym{adc}{ADC}{Analog to Digital Converter}
\newacronym{aerpaw}{AERPAW}{Aerial Experimentation and Research Platform for Advanced Wireless}
\newacronym{ai}{AI}{Artificial Intelligence}
\newacronym{aimd}{AIMD}{Additive Increase Multiplicative Decrease}
\newacronym{am}{AM}{Acknowledged Mode}
\newacronym{amc}{AMC}{Adaptive Modulation and Coding}
\newacronym{amf}{AMF}{Access and Mobility Management Function}
\newacronym{aops}{AOPS}{Adaptive Order Prediction Scheduling}
\newacronym{api}{API}{Application Programming Interface}
\newacronym{apn}{APN}{Access Point Name}
\newacronym{ap}{AP}{Application Protocol}
\newacronym{aqm}{AQM}{Active Queue Management}
\newacronym{ausf}{AUSF}{Authentication Server Function}
\newacronym{avc}{AVC}{Advanced Video Coding}
\newacronym{awgn}{AGWN}{Additive White Gaussian Noise}
\newacronym{balia}{BALIA}{Balanced Link Adaptation Algorithm}
\newacronym{bbu}{BBU}{Base Band Unit}
\newacronym{bdp}{BDP}{Bandwidth-Delay Product}
\newacronym{ber}{BER}{Bit Error Rate}
\newacronym{bf}{BF}{Beamforming}
\newacronym{bler}{BLER}{Block Error Rate}
\newacronym{brr}{BRR}{Bayesian Ridge Regressor}
\newacronym{bs}{BS}{Base Station}
\newacronym{bsr}{BSR}{Buffer Status Report}
\newacronym{bss}{BSS}{Business Support System}
\newacronym{ca}{CA}{Carrier Aggregation}
\newacronym{caas}{CaaS}{Connectivity-as-a-Service}
\newacronym{cav}{CAV}{Connected and Autonoums Vehicle}
\newacronym{cb}{CB}{Code Block}
\newacronym{cc}{CC}{Congestion Control}
\newacronym{ccid}{CCID}{Congestion Control ID}
\newacronym{cco}{CC}{Carrier Component}
\newacronym{cd}{CD}{Continuous Delivery}
\newacronym{cdd}{CDD}{Cyclic Delay Diversity}
\newacronym{cdf}{CDF}{Cumulative Distribution Function}
\newacronym{cdn}{CDN}{Content Distribution Network}
\newacronym{cli}{CLI}{Command-line Interface}
\newacronym{cn}{CN}{Core Network}
\newacronym{codel}{CoDel}{Controlled Delay Management}
\newacronym{comac}{COMAC}{Converged Multi-Access and Core}
\newacronym{cord}{CORD}{Central Office Re-architected as a Datacenter}
\newacronym{cornet}{CORNET}{COgnitive Radio NETwork}
\newacronym{cosmos}{COSMOS}{Cloud Enhanced Open Software Defined Mobile Wireless Testbed for City-Scale Deployment}
\newacronym{cots}{COTS}{Commercial Off-the-Shelf}
\newacronym{cp}{CP}{Control Plane}
\newacronym{cyp}{CP}{Cyclic Prefix}
\newacronym{up}{UP}{User Plane}
\newacronym{cpu}{CPU}{Central Processing Unit}
\newacronym{cqi}{CQI}{Channel Quality Information}
\newacronym{cr}{CR}{Cognitive Radio}
\newacronym{cran}{CRAN}{Cloud \gls{ran}}
\newacronym{crs}{CRS}{Cell Reference Signal}
\newacronym{csi}{CSI}{Channel State Information}
\newacronym{csirs}{CSI-RS}{Channel State Information - Reference Signal}
\newacronym{cu}{CU}{Central Unit}
\newacronym{d2tcp}{D$^2$TCP}{Deadline-aware Data center TCP}
\newacronym{d3}{D$^3$}{Deadline-Driven Delivery}
\newacronym{dac}{DAC}{Digital to Analog Converter}
\newacronym{dag}{DAG}{Directed Acyclic Graph}
\newacronym{das}{DAS}{Distributed Antenna System}
\newacronym{dash}{DASH}{Dynamic Adaptive Streaming over HTTP}
\newacronym{dc}{DC}{Dual Connectivity}
\newacronym{dccp}{DCCP}{Datagram Congestion Control Protocol}
\newacronym{dce}{DCE}{Direct Code Execution}
\newacronym{dci}{DCI}{Downlink Control Information}
\newacronym{dctcp}{DCTCP}{Data Center TCP}
\newacronym{dl}{DL}{Downlink}
\newacronym{dmr}{DMR}{Deadline Miss Ratio}
\newacronym{dmrs}{DMRS}{DeModulation Reference Signal}
\newacronym{drlcc}{DRL-CC}{Deep Reinforcement Learning Congestion Control}
\newacronym{dsrc}{DSRC}
{dedicated short-range communications}
\newacronym{d2d}{D2D}{device-to-device}
\newacronym{drs}{DRS}{Discovery Reference Signal}
\newacronym{du}{DU}{Distributed Unit}
\newacronym{e2e}{E2E}{end-to-end}
\newacronym{earfcn}{EARFCN}{E-UTRA Absolute Radio Frequency Channel Number}
\newacronym{ecaas}{ECaaS}{Edge-Cloud-as-a-Service}
\newacronym{ecn}{ECN}{Explicit Congestion Notification}
\newacronym{edf}{EDF}{Earliest Deadline First}
\newacronym{embb}{eMBB}{Enhanced Mobile Broadband}
\newacronym{empower}{EMPOWER}{EMpowering transatlantic PlatfOrms for advanced WirEless Research}
\newacronym{enb}{eNB}{evolved Node Base}
\newacronym{endc}{EN-DC}{E-UTRAN-\gls{nr} \gls{dc}}
\newacronym{epc}{EPC}{Evolved Packet Core}
\newacronym{eps}{EPS}{Evolved Packet System}
\newacronym{es}{ES}{Edge Server}
\newacronym{etsi}{ETSI}{European Telecommunications Standards Institute}
\newacronym[firstplural=Estimated Times of Arrival (ETAs)]{eta}{ETA}{Estimated Time of Arrival}
\newacronym{eutran}{E-UTRAN}{Evolved Universal Terrestrial Access Network}
\newacronym{faas}{FaaS}{Function-as-a-Service}
\newacronym{fapi}{FAPI}{Functional Application Platform Interface}
\newacronym{fdd}{FDD}{Frequency Division Duplexing}
\newacronym{fdm}{FDM}{Frequency Division Multiplexing}
\newacronym{fdma}{FDMA}{Frequency Division Multiple Access}
\newacronym{fed4fire}{FED4FIRE+}{Federation 4 Future Internet Research and Experimentation Plus}
\newacronym{fir}{FIR}{Finite Impulse Response}
\newacronym{fit}{FIT}{Future \acrlong{iot}}
\newacronym{fpga}{FPGA}{Field Programmable Gate Array}
\newacronym{fr2}{FR2}{Frequency Range 2}
\newacronym{fr1}{FR1}{Frequency Range 1}
\newacronym{fs}{FS}{Fast Switching}
\newacronym{fscc}{FSCC}{Flow Sharing Congestion Control}
\newacronym{ftp}{FTP}{File Transfer Protocol}
\newacronym{fw}{FW}{Flow Window}
\newacronym{ge}{GE}{Gaussian Elimination}
\newacronym{gnb}{gNB}{Next Generation Node Base}
\newacronym{gop}{GOP}{Group of Pictures}
\newacronym{gpr}{GPR}{Gaussian Process Regressor}
\newacronym{gpu}{GPU}{Graphics Processing Unit}
\newacronym{gtp}{GTP}{GPRS Tunneling Protocol}
\newacronym{gtpc}{GTP-C}{GPRS Tunnelling Protocol Control Plane}
\newacronym{gtpu}{GTP-U}{GPRS Tunnelling Protocol User Plane}
\newacronym{gtpv2c}{GTPv2-C}{\gls{gtp} v2 - Control}
\newacronym{gw}{GW}{Gateway}
\newacronym{harq}{HARQ}{Hybrid Automatic Repeat reQuest}
\newacronym{hetnet}{HetNet}{Heterogeneous Network}
\newacronym{hh}{HH}{Hard Handover}
\newacronym{hol}{HOL}{Head-of-Line}
\newacronym{hqf}{HQF}{Highest-quality-first}
\newacronym{hss}{HSS}{Home Subscription Server}
\newacronym{http}{HTTP}{HyperText Transfer Protocol}
\newacronym{ia}{IA}{Initial Access}
\newacronym{iab}{IAB}{Integrated Access and Backhaul}
\newacronym{ic}{IC}{Incident Command}
\newacronym{ietf}{IETF}{Internet Engineering Task Force}
\newacronym{imsi}{IMSI}{International Mobile Subscriber Identity}
\newacronym{imt}{IMT}{International Mobile Telecommunication}
\newacronym{iot}{IoT}{Internet of Things}
\newacronym{ip}{IP}{Internet Protocol}
\newacronym{itu}{ITU}{International Telecommunication Union}
\newacronym{kpi}{KPI}{Key Performance Indicator}
\newacronym{kpm}{KPM}{Key Performance Measurement}
\newacronym{kvm}{KVM}{Kernel-based Virtual Machine}
\newacronym{los}{LoS}{Line of Sight}
\newacronym{lsm}{LSM}{Link-to-System Mapping}
\newacronym{lstm}{LSTM}{Long Short Term Memory}
\newacronym{lte}{LTE}{Long Term Evolution}
\newacronym{lxc}{LXC}{Linux Container}
\newacronym{m2m}{M2M}{Machine to Machine}
\newacronym{mac}{MAC}{Medium Access Control}
\newacronym{manet}{MANET}{Mobile Ad Hoc Network}
\newacronym{mano}{MANO}{Management and Orchestration}
\newacronym{mc}{MC}{Multi-Connectivity}
\newacronym{mcc}{MCC}{Mobile Cloud Computing}
\newacronym{mchem}{MCHEM}{Massive Channel Emulator}
\newacronym{mcs}{MCS}{Modulation and Coding Scheme}
\newacronym{mec2}{MEC}{Multi-access Edge Computing}
\newacronym{mec}{MEC}{Mobile Edge Computing}
\newacronym{mfc}{MFC}{Mobile Fog Computing}
\newacronym{mgen}{MGEN}{Multi-Generator}
\newacronym{mi}{MI}{Mutual Information}
\newacronym{mib}{MIB}{Master Information Block}
\newacronym{miesm}{MIESM}{Mutual Information Based Effective SINR}
\newacronym{mimo}{MIMO}{Multiple Input, Multiple Output}
\newacronym{ml}{ML}{Machine Learning}
\newacronym{mlr}{MLR}{Maximum-local-rate}
\newacronym[plural=\gls{mme}s,firstplural=Mobility Management Entities (MMEs)]{mme}{MME}{Mobility Management Entity}
\newacronym{mmtc}{mMTC}{Massive Machine-Type Communications}
\newacronym{mmwave}{mmWave}{millimeter wave}
\newacronym{mpdccp}{MP-DCCP}{Multipath Datagram Congestion Control Protocol}
\newacronym{mptcp}{MPTCP}{Multipath TCP}
\newacronym{mr}{MR}{Maximum Rate}
\newacronym{mrdc}{MR-DC}{Multi \gls{rat} \gls{dc}}
\newacronym{mse}{MSE}{Mean Square Error}
\newacronym{mss}{MSS}{Maximum Segment Size}
\newacronym{mt}{MT}{Mobile Termination}
\newacronym{mtd}{MTD}{Machine-Type Device}
\newacronym{mtu}{MTU}{Maximum Transmission Unit}
\newacronym{mumimo}{MU-MIMO}{Multi-user \gls{mimo}}
\newacronym{mvno}{MVNO}{Mobile Virtual Network Operator}
\newacronym{nalu}{NALU}{Network Abstraction Layer Unit}
\newacronym{nas}{NAS}{Network Attached Storage}
\newacronym{nat}{NAT}{Network Address Translation}
\newacronym{nbiot}{NB-IoT}{Narrow Band IoT}
\newacronym{nfv}{NFV}{Network Function Virtualization}
\newacronym{nfvi}{NFVI}{Network Function Virtualization Infrastructure}
\newacronym{ni}{NI}{Network Interfaces}
\newacronym{nic}{NIC}{Network Interface Card}
\newacronym{now}{NOW}{Non Overlapping Window}
\newacronym{nsm}{NSM}{Network Service Mesh}
\newacronym{nr}{NR}{New Radio}
\newacronym{nrf}{NRF}{Network Repository Function}
\newacronym{nsa}{NSA}{Non Stand Alone}
\newacronym{nse}{NSE}{Network Slicing Engine}
\newacronym{nssf}{NSSF}{Network Slice Selection Function}
\newacronym{o2i}{O2I}{Outdoor to Indoor}
\newacronym{oai}{OAI}{OpenAirInterface}
\newacronym{oaicn}{OAI-CN}{\gls{oai} \acrlong{cn}}
\newacronym{oairan}{OAI-RAN}{\acrlong{oai} \acrlong{ran}}
\newacronym{oam}{OAM}{Operations, Administration and Maintenance}
\newacronym{ofdm}{OFDM}{Orthogonal Frequency Division Multiplexing}
\newacronym{olia}{OLIA}{Opportunistic Linked Increase Algorithm}
\newacronym{omec}{OMEC}{Open Mobile Evolved Core}
\newacronym{onap}{ONAP}{Open Network Automation Platform}
\newacronym{onf}{ONF}{Open Networking Foundation}
\newacronym{onos}{ONOS}{Open Networking Operating System}
\newacronym{oom}{OOM}{\gls{onap} Operations Manager}
\newacronym{opnfv}{OPNFV}{Open Platform for \gls{nfv}}
\newacronym{oran}{O-RAN}{Open \gls{ran}}
\newacronym{orbit}{ORBIT}{Open-Access Research Testbed for Next-Generation Wireless Networks}
\newacronym{os}{OS}{Operating System}
\newacronym{oss}{OSS}{Operations Support System}
\newacronym{pa}{PA}{Position-aware}
\newacronym{pase}{PASE}{Prioritization, Arbitration, and Self-adjusting Endpoints}
\newacronym{pawr}{PAWR}{Platforms for Advanced Wireless Research}
\newacronym{pbch}{PBCH}{Physical Broadcast Channel}
\newacronym{pcef}{PCEF}{Policy and Charging Enforcement Function}
\newacronym{pcfich}{PCFICH}{Physical Control Format Indicator Channel}
\newacronym{pcrf}{PCRF}{Policy and Charging Rules Function}
\newacronym{pdcch}{PDCCH}{Physical Downlink Control Channel}
\newacronym{pdcp}{PDCP}{Packet Data Convergence Protocol}
\newacronym{pdsch}{PDSCH}{Physical Downlink Shared Channel}
\newacronym{pdu}{PDU}{Packet Data Unit}
\newacronym{pf}{PF}{Proportional Fair}
\newacronym{pgw}{PGW}{Packet Gateway}
\newacronym{phich}{PHICH}{Physical Hybrid ARQ Indicator Channel}
\newacronym{phy}{PHY}{Physical}
\newacronym{pmch}{PMCH}{Physical Multicast Channel}
\newacronym{pmi}{PMI}{Precoding Matrix Indicators}
\newacronym{powder}{POWDER}{Platform for Open Wireless Data-driven Experimental Research}
\newacronym{ppo}{PPO}{Proximal Policy Optimization}
\newacronym{ppp}{PPP}{Poisson Point Process}
\newacronym{prach}{PRACH}{Physical Random Access Channel}
\newacronym{prb}{PRB}{Physical Resource Block}
\newacronym{psnr}{PSNR}{Peak Signal to Noise Ratio}
\newacronym{pss}{PSS}{Primary Synchronization Signal}
\newacronym{pucch}{PUCCH}{Physical Uplink Control Channel}
\newacronym{pusch}{PUSCH}{Physical Uplink Shared Channel}
\newacronym{qam}{QAM}{Quadrature Amplitude Modulation}
\newacronym{qci}{QCI}{\gls{qos} Class Identifier}
\newacronym{qoe}{QoE}{Quality of Experience}
\newacronym{qos}{QoS}{Quality of Service}
\newacronym{quic}{QUIC}{Quick UDP Internet Connections}
\newacronym{ra}{RA}{Resouces Allocation}
\newacronym{rach}{RACH}{Random Access Channel}
\newacronym{ran}{RAN}{Radio Access Network}
\newacronym[firstplural=Radio Access Technologies (RATs)]{rat}{RAT}{Radio Access Technology}
\newacronym{rbg}{RBG}{Resource Block Group}
\newacronym{rcn}{RCN}{Research Coordination Network}
\newacronym{rc}{RC}{RAN Control}
\newacronym{rec}{REC}{Radio Edge Cloud}
\newacronym{red}{RED}{Random Early Detection}
\newacronym{renew}{RENEW}{Reconfigurable Eco-system for Next-generation End-to-end Wireless}
\newacronym{rf}{RF}{Radio Frequency}
\newacronym{rfc}{RFC}{Request for Comments}
\newacronym{rfr}{RFR}{Random Forest Regressor}
\newacronym{ric}{RIC}{\gls{ran} Intelligent Controller}
\newacronym{rlc}{RLC}{Radio Link Control}
\newacronym{rlf}{RLF}{Radio Link Failure}
\newacronym{rlnc}{RLNC}{Random Linear Network Coding}
\newacronym{rmr}{RMR}{RIC Message Router}
\newacronym{rmse}{RMSE}{Root Mean Squared Error}
\newacronym{rnis}{RNIS}{Radio Network Information Service}
\newacronym{rr}{RR}{Round Robin}
\newacronym{rrc}{RRC}{Radio Resource Control}
\newacronym{rrm}{RRM}{Radio Resource Management}
\newacronym{rru}{RRU}{Remote Radio Unit}
\newacronym{rs}{RS}{Remote Server}
\newacronym{rsrp}{RSRP}{Reference Signal Received Power}
\newacronym{rsrq}{RSRQ}{Reference Signal Received Quality}
\newacronym{rss}{RSS}{Received Signal Strength}
\newacronym{rssi}{RSSI}{Received Signal Strength Indicator}
\newacronym{rtt}{RTT}{Round Trip Time}
\newacronym{ru}{RU}{Radio Unit}
\newacronym{rus}{RSU}{Road Side Unit}
\newacronym{rw}{RW}{Receive Window}
\newacronym{rx}{RX}{Receiver}
\newacronym{s1ap}{S1AP}{S1 Application Protocol}
\newacronym{sa}{SA}{standalone}
\newacronym{sack}{SACK}{Selective Acknowledgment}
\newacronym{sap}{SAP}{Service Access Point}
\newacronym{sc2}{SC2}{Spectrum Collaboration Challenge}
\newacronym{scef}{SCEF}{Service Capability Exposure Function}
\newacronym{sch}{SCH}{Secondary Cell Handover}
\newacronym{scoot}{SCOOT}{Split Cycle Offset Optimization Technique}
\newacronym{sctp}{SCTP}{Stream Control Transmission Protocol}
\newacronym{sdap}{SDAP}{Service Data Adaptation Protocol}
\newacronym{sdk}{SDK}{Software Development Kit}
\newacronym{sdm}{SDM}{Space Division Multiplexing}
\newacronym{sdma}{SDMA}{Spatial Division Multiple Access}
\newacronym{sdn}{SDN}{Software-defined Networking}
\newacronym{sdr}{SDR}{Software-defined Radio}
\newacronym{seba}{SEBA}{SDN-Enabled Broadband Access}
\newacronym{sgsn}{SGSN}{Serving GPRS Support Node}
\newacronym{sgw}{SGW}{Service Gateway}
\newacronym{si}{SI}{Study Item}
\newacronym{sib}{SIB}{Secondary Information Block}
\newacronym{sinr}{SINR}{Signal to Interference plus Noise Ratio}
\newacronym{sip}{SIP}{Session Initiation Protocol}
\newacronym{siso}{SISO}{Single Input, Single Output}
\newacronym{sla}{SLA}{Service Level Agreement}
\newacronym{sm}{SM}{Service Model}
\newacronym{smo}{SMO}{Service Management and Orchestration}
\newacronym{smsgmsc}{SMS-GMSC}{\gls{sms}-Gateway}
\newacronym{snr}{SNR}{Signal-to-Noise-Ratio}
\newacronym{son}{SON}{Self-Organizing Network}
\newacronym{sptcp}{SPTCP}{Single Path TCP}
\newacronym{srb}{SRB}{Service Radio Bearer}
\newacronym{srn}{SRN}{Standard Radio Node}
\newacronym{srs}{SRS}{Sounding Reference Signal}
\newacronym{ss}{SS}{Synchronization Signal}
\newacronym{sss}{SSS}{Secondary Synchronization Signal}
\newacronym{st}{ST}{Spanning Tree}
\newacronym{svc}{SVC}{Scalable Video Coding}
\newacronym{tb}{TB}{Transport Block}
\newacronym{tcp}{TCP}{Transmission Control Protocol}
\newacronym{tdd}{TDD}{Time Division Duplexing}
\newacronym{tdm}{TDM}{Time Division Multiplexing}
\newacronym{tdma}{TDMA}{Time Division Multiple Access}
\newacronym{tfl}{TfL}{Transport for London}
\newacronym{tfrc}{TFRC}{TCP-Friendly Rate Control}
\newacronym{tft}{TFT}{Traffic Flow Template}
\newacronym{tgen}{TGEN}{Traffic Generator}
\newacronym{tip}{TIP}{Telecom Infra Project}
\newacronym{tm}{TM}{Transparent Mode}
\newacronym{to}{TO}{Telco Operator}
\newacronym{tr}{TR}{Technical Report}
\newacronym{trp}{TRP}{Transmitter Receiver Pair}
\newacronym{ts}{TS}{Technical Specification}
\newacronym{tti}{TTI}{Transmission Time Interval}
\newacronym{ttt}{TTT}{Time-to-Trigger}
\newacronym{tx}{TX}{Transmitter}
\newacronym{uas}{UAS}{Unmanned Aerial System}
\newacronym{uav}{UAV}{Unmanned Aerial Vehicle}
\newacronym{udm}{UDM}{Unified Data Management}
\newacronym{udp}{UDP}{User Datagram Protocol}
\newacronym{udr}{UDR}{Unified Data Repository}
\newacronym{ue}{UE}{User Equipment}
\newacronym{uhd}{UHD}{\gls{usrp} Hardware Driver}
\newacronym{ul}{UL}{Uplink}
\newacronym{um}{UM}{Unacknowledged Mode}
\newacronym{uml}{UML}{Unified Modeling Language}
\newacronym{upa}{UPA}{Uniform Planar Array}
\newacronym{upf}{UPF}{User Plane Function}
\newacronym{urllc}{URLLC}{Ultra Reliable and Low Latency Communications}
\newacronym{usa}{U.S.}{United States}
\newacronym{usim}{USIM}{Universal Subscriber Identity Module}
\newacronym{usrp}{USRP}{Universal Software Radio Peripheral}
\newacronym{utc}{UTC}{Urban Traffic Control}
\newacronym{vim}{VIM}{Virtualization Infrastructure Manager}
\newacronym{vm}{VM}{Virtual Machine}
\newacronym{vnf}{VNF}{Virtual Network Function}
\newacronym{volte}{VoLTE}{Voice over \gls{lte}}
\newacronym{voltha}{VOLTHA}{Virtual OLT HArdware Abstraction}
\newacronym{vr}{VR}{Virtual Reality}
\newacronym{vran}{vRAN}{Virtualized \gls{ran}}
\newacronym{vss}{VSS}{Video Streaming Server}
\newacronym{v2x}{V2X}{vehicle-to-everything}
\newacronym{v2i}{V2I}{vehicle-to-infrastructure}
\newacronym{v2v}{V2V}{vehicle-to-vehicle}
\newacronym{v2n}{V2N}{vehicle-to-network}
\newacronym{wbf}{WBF}{Wired Bias Function}
\newacronym{wf}{WF}{Waterfilling}
\newacronym{wg}{WG}{Working Group}
\newacronym{wlan}{WLAN}{Wireless Local Area Network}
\newacronym{osm}{OSM}{Open Source \gls{nfv} Management and Orchestration}
\newacronym{pnf}{PNF}{Physical Network Function}
\newacronym{drl}{DRL}{Deep Reinforcement Learning}
\newacronym{mtc}{MTC}{Machine-type Communications}
\newacronym{osc}{OSC}{O-RAN Software Community}
\newacronym{mns}{MnS}{Management Services}
\newacronym{ves}{VES}{\gls{vnf} Event Stream}
\newacronym{ei}{EI}{Enrichment Information}
\newacronym{fh}{FH}{Fronthaul}
\newacronym{fft}{FFT}{Fast Fourier Transform}
\newacronym{laa}{LAA}{Licensed-Assisted Access}
\newacronym{plfs}{PLFS}{Physical Layer Frequency Signals}
\newacronym{ptp}{PTP}{Precision Time Protocol}
\newacronym{lidar}{LiDAR}{Light Detection And Ranging}
\newacronym{dem}{DEM}{Digital Elevation Model}
\newacronym{dtm}{DEM}{Digital Terrain Model}
\newacronym{dsm}{DEM}{Digital Surface Models}
\newacronym{ota}{OTA}{Over-The-Air}
\newacronym{ns}{NS}{Network Slicing}
\newacronym{ne}{NE}{Nash Equilibrium}
\newacronym{hf}{HF}{High Frequency}
\newacronym{noma}{NOMA}{Non-Orthogonal Multiple Access}
\newacronym{sre}{SRE}{Smart Radio Environment}
\newacronym{ris}{RIS}{Reconfigurable Intelligent Surface}
\newacronym{inp}{InP}{Infrastructure Provider}
\newacronym{smf}{SMF}{Slicing Magangement Framework}
\newacronym{nsn}{NSN}{Network Slicing Negotiation}
\newacronym{sms}{SMS}{Slicing MAC Scheduler}
\newacronym{brd}{BRD}{Best Response Dynamics}
\newacronym{dssbr}{DSSBR}{Double Step Smoothed Best Response}
\newacronym{poa}{PoA}{Price of Anarchy}
\newacronym{pos}{PoS}{Price of Stability}
\newacronym{milp}{MILP}{Mixed Integer-Linear Program}
\newacronym{pod}{PoD}{Price of DSSBR}
\newacronym{roc}{ROC}{Radio Overload Control}
\newacronym{ciot}{cIoT}{critical Internet of Things}
\newacronym{embbpr}{eMBB Pr.}{enhanced Mobile BroadBand Premium}
\newacronym{sps}{SPS}{Semi-persistent Scheduling}
\newacronym{cg}{CG}{Configured Grant}
\newacronym{embbbs}{eMBB Bs.}{enhanced Mobile BroadBand Basic}
\newacronym{en}{EN}{Edge Node}
\newacronym{ec}{EC}{Edge Computing}
\newacronym{sp}{SP}{Service Provider}
\newacronym{me}{ME}{Market Equilibrium}
\newacronym{so}{SO}{Social Optimum}
\newacronym{wso}{WSO}{Weighted Social Optimum}
\newacronym{ps}{PS}{Proportional Sharing}
\newacronym{eg}{EG}{Eisenberg-Gale program}
\newacronym{pe}{PE}{Pareto Efficiency}
\newacronym{nsw}{NSW}{Nash Social Welfare}
\newacronym{ef}{EF}{Envy-Freeness}
\newacronym{sub6}{sub-$6$GHz}{Below $6\,$GHz}
\newacronym{ncr}{NCR}{Network-Controlled Repeater}
\newacronym{nlos}{NLoS}{Non-Line of Sight}
\newacronym{src}{SRC}{Smart Radio Connection}
\newacronym{srd}{SRD}{Smart Radio Device}
\newacronym{cs}{CS}{Candidate Site}
\newacronym{tp}{TP}{Test Point}
\newacronym{fov}{FoV}{Field of View}
\newacronym{nrric}{near-RT RIC}{Near Real-time {RAN} Intelligent Controller}
\newacronym{e2ap}{E2AP}{E2 Application Protocol}
\newacronym{e2sm}{E2SM}{E2 Service Model}
\newacronym{nrtric}{non-RT RIC}{Non-Real-Time {RIC}}
\newacronym{itti}{ITTI}{Inter-task Interface}
\newacronym{bap}{BAP}{Backhaul Adaptation Protocol}
\newacronym{iabest}{IABEST}{Integrated Access and Backhaul Experimental large-Scale Tetbed}
\newacronym{teid}{TEID}{Tunnel Endpoint Identifier}
\newacronym{dlsch}{DL-SCH}{Downlink Shared Channel }
\newacronym{ulsch}{UL-SCH}{Uplink Shared Channel }
\newacronym{rsu}{RSU}{Road Side Unit}
\newacronym{its}{ITS}{Intelligent Transportation Systems}
\newacronym{vanet}{VANET}{Vehicular Ad-hoc Network}
\newacronym{dt}{DT}{Digital Twin}
\newacronym{ecc}{ECC}{Edge Computing Cluster}
\newacronym{fig}{Fig.}{Figure}

\def\BibTeX{{\rm B\kern-.05em{\sc i\kern-.025em b}\kern-.08em
    T\kern-.1667em\lower.7ex\hbox{E}\kern-.125emX}}

\title{Advancing O-RAN to Facilitate Intelligence in V2X}

\author{Eugenio Moro,\thanks{The authors are with the Dipartimento di Elettronica Informazione e Bioingegneria, Politecnico di Milano, Milano, Italy, (\mbox{e-mails}: \{name.surname\}@polimi.it).\\ This article was supported by the European Union under the Italian National Recovery and Resilience Plan (NRRP) of NextGenerationEU, partnership on “Telecommunications of the Future” (PE00000001 - program “RESTART”, Structural Project 6GWINET)} Francesco~Linsalata, Maurizio~Magarini, Umberto~Spagnolini, and~Antonio~Capone}

\begin{document}
\maketitle

\begin{abstract}

Vehicular communications integrated with the \gls{ran} are envisioned as a breakthrough application for the \gls{6g} cellular systems. However, traditional \glspl{ran} lack the flexibility to enable sophisticated control mechanisms that are demanded by the strict performance requirements of the \gls{v2x} environment.
In contrast, the features of \gls{oran} can be exploited to support advanced use cases, as its core paradigms represent an ideal framework for orchestrating vehicular communication. 
Although the high potential stemming from their integration can be easily seen and recognized, the effective combination of the two ecosystems is an open issue. Conceptual and architectural advances are required for \gls{oran} to be capable of facilitating network intelligence in \gls{v2x}.
This article pioneers the integration of the two strategies for seamlessly incorporating \gls{v2x} control within \gls{oran}'s ecosystem.
First, an enabling architecture that tightly integrates \gls{v2x} and \gls{oran} is proposed and discussed. Then, a set of key \gls{v2x} challenges is identified, and \gls{oran}-based solutions are proposed, paired with extensive numerical analysis to support their effectiveness.
Results showcase the superior performance of such an approach in terms of raw throughput, network resilience, and control overhead. Finally, these results validate the proposed enabling architecture and confirm the potential of \gls{oran} in support of \gls{v2x} communications.
\end{abstract}

\begin{IEEEkeywords}
O-RAN, V2X, 6G, network intelligence, dynamic control
\end{IEEEkeywords}

\section{Introduction}

The current \gls{ran} paradigm limits network intelligence due to network components - mainly \gls{bs} - being operated as monolithic and inflexible black-boxes~\cite{polese2022understanding}. To address this limitation and bridge the gap between real-world \gls{ran} deployments and cutting-edge network intelligence, a consortium of vendors, operators, and research institutions have proposed \gls{oran} as an architectural overhaul of \glspl{ran}.

\gls{oran} is a disaggregated, open architecture that separates hardware and software components, enabling interoperability, modularity, and flexibility \cite{polese2022understanding}. A major innovation of \gls{oran} is the \glspl{ric}, softwarized control loops for data collection and dynamic control, implemented as micro-services over large-scale, heterogeneous \gls{ran} deployments.
\gls{oran}-based control solutions have optimized various aspects of \gls{5g} systems, demonstrating the architecture's transformative impact. However, \gls{oran} currently supports only traditional \gls{5g} deployments, with network components limited to \glspl{bs} and \glspl{ue}.
Advancing \gls{oran} to support emerging 6G use cases, where network intelligence is crucial, is an essential research endeavour~\cite{moro2023toward}.

In this work, we argue that \glspl{cav} exploiting high data-rate links represents one of these fundamental use cases, and we propose our vision on the matter. Indeed, vehicular communication will likely be a key driving force in the future \gls{6g} wireless networks, enabling advanced vehicular mobility paradigms such as autonomous driving, coordinated sensing, and enhanced navigation. 

At the core of vehicular communications lies \gls{v2x}, a communication technology that facilitates the interconnection among vehicles and infrastructure. \Gls{v2x} realizes direct and multi-hops links among \glspl{cav}, namely \gls{v2v} or Sidelink communications. Direct vehicular links reduce the network infrastructure involvement, facilitating communication even in out-of-coverage areas and considerably decreasing latency \cite{6GV2Xchallenges}. Moreover, since most of the \glspl{cav} use cases require high data rates, \gls{v2v} will make use of higher carrier frequencies, such as \gls{mmwave} - currently standardized in \gls{5g} as \gls{fr2} - or sub-THz, with the introduction of beam-based highly directive communication to counteract the considerable pathloss that characterizes propagation in these bands \cite{6GV2Xchallenges, Gozalvez8887840}. 

The unique challenges posed by \gls{cav} scenario, the dynamic nature of the vehicular environment, the harsh propagation conditions at high frequencies, as well as the hybrid nature of \gls{v2x} necessitate the development of sophisticated control mechanisms to ensure the success of this disruptive technology. Albeit currently being limited to traditional \gls{ran} deployments support, \gls{oran} represents the ideal candidate to enable management and orchestration in the challenging scenario mentioned above.

However, the opportunities of enabling network intelligence for \glspl{cav} through \gls{oran} have not sufficiently explored in the literature. In particular, advancing \gls{oran} to fully support \gls{v2x} orchestration down to the control of the individual \gls{cav}'s communication stack has never been discussed.

A next-generation \gls{oran} architecture is proposed, tightly integrating \gls{v2x} within \gls{oran} concepts for the first time. By extending \gls{oran} interfaces, we support additional \gls{v2x} network components managed by \gls{oran} control loops. Notably, the communication stack of connected vehicles becomes part of the \gls{oran} architecture. This results in a comprehensive vehicular communication solution where \gls{oran} \glspl{ric} orchestrate a hybrid network, connecting vehicles to the \gls{ran} and to each other. Reliable low-frequency \glspl{ran} (e.g., 5G \gls{fr1}) support a control plane for \gls{oran} messages between vehicles and the \gls{ric}. In turn, the \gls{ric} manages high-frequency \gls{v2v} links to create a high-performance data plane, a \gls{vanet}, for autonomous driving, infotainment, and other applications.

After having defined the supporting architecture, the focus shifts on 
%
%integrating next-generation \gls{oran} with \gls{v2x} communications, 
highlighting the challenges and opportunities to enhance the performance, reliability, and adaptability of \gls{ran}-based vehicular communications through \gls{oran}. We identify the following key \gls{v2x} challenges to be tackled by \gls{oran}: beam selection and management, radio resource management, \gls{v2v} connectivity, \gls{v2x} digital twin creation and optimal use.
We explore how \gls{oran}-based solutions can address each of these challenges, presenting potential research directions.

While state-of-the-art solutions exist for every identified challenge, none immediately apply to a real-world \gls{v2x} deployment. On the other hand, \gls{oran}-based solutions are immediately feasible by definition. A fair comparison can only be obtained against the existing standard specifications. 
For this reason, all of the selected vital challenges are paired with a dedicated numerical analysis where a demonstrative \gls{oran}-based solution is compared to standard-based mechanisms.
In detail, we evaluate \gls{oran} micro-services (xApps) in a bespoke \gls{v2x} simulation framework~\cite{transaction_oran_v2x}, showcasing their capabilities and benefits in solving real-world \gls{v2x} problems and confirming the superior performance.
\begin{figure*}[!t]
	\centering \includegraphics[width=0.98\textwidth]{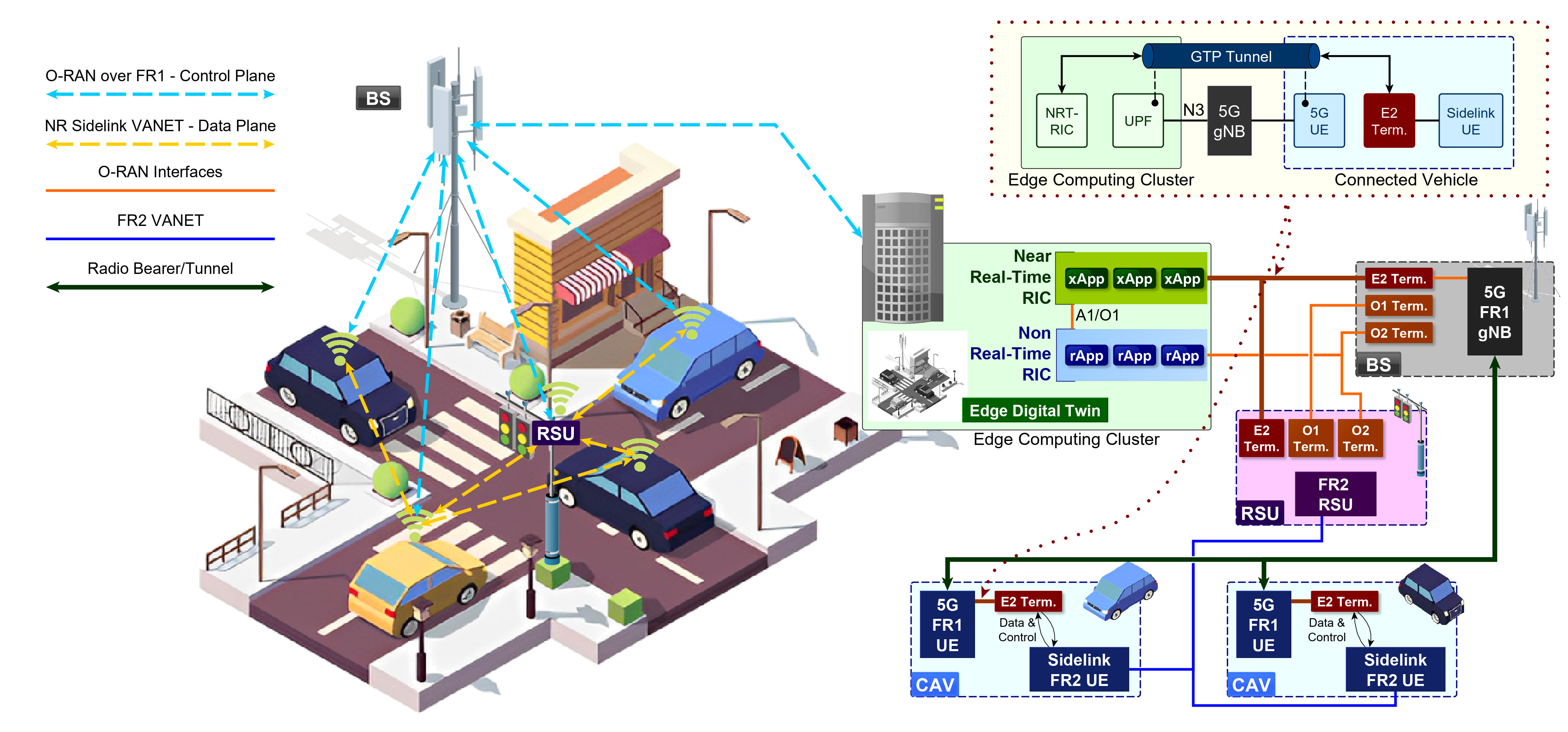}
	\caption{Next-generation \gls{oran} architecture details.}
	\label{fig:ric}
\end{figure*}
\section{A next-gen \gls{oran} architecture for \gls{v2x}}
\label{sec:architecture}
The \gls{oran} architecture features the possibility of applying centralized control to the \gls{ran} through the \glspl{ric}. These functional components can implement arbitrary data collection and control logic by communicating with the network infrastructure (i.e., \glspl{bs}) thanks to open and standardized interfaces. In particular, \gls{oran} introduced a \gls{nrric}, which operates on a $1\,$ms to $1\,$s time scale and, thus, it is capable of operating under stringent \gls{v2x} latency requirements. Arbitrary data collection and control mechanisms are then implemented through xApps, namely network microservices that run on the data collection and control primitives exposed by the \gls{nrric}. Additionally, \gls{oran} has standardized the \gls{nrtric}: a centralized control loop operating on a slower time scale but with broader network visibility. \Glspl{nrtric} enable large-scale orchestration and policing mechanisms implemented as network applications called rApps. When applied to \gls{v2x}, these two control loops can potentially unlock significant optimization and orchestration gains with respect to the current architecture. 

Due to the peculiarities of the V2X environment, some extensions to the \gls{oran} architecture are required such that all of these solutions can be practically realized. We now focus on this matter, proposing our vision of an enabling architecture where \gls{oran} and \gls{v2x} are integrated to unlock the opportunities above. As shown in Figure~\ref{fig:ric} for the case of using \gls{5g} as the \gls{rat}, a typical \gls{oran} deployment includes a \gls{nrtric} and a \gls{nrric} embodied as software components in the \gls{ecc}~\cite{polese2022understanding}. A \gls{dt} could also be deployed inside the same \gls{ecc}, allowing for undisturbed data exchanges with the \glspl{ric}. \Gls{oran} apps running on top of both \glspl{ric} communicate with the network infrastructure through open interfaces: the E2 interface for the \gls{nrtric} and the O1 and O2 interfaces for the \gls{nrric}.

Consequently, designing an \gls{oran} empowered \gls{v2x} system requires all the \gls{v2x} devices to be reachable by the \glspl{ric} through these interfaces. \Glspl{bs} are normally equipped with these interface terminations, as shown in Figure~\ref{fig:ric}; thus, no architectural modifications are required. However, the interface definitions will likely require extensions to support the specifics of data collection and control required by vehicular use cases. \Glspl{rsu} represent an important architectural component of \gls{v2x} networks. These communication devices provide \gls{v2i} connectivity to nearby vehicles and can act as relays, enhancing \gls{v2v} connections~\cite{heo2019rsu}. \Glspl{rsu} are currently not supported by \gls{oran} specifications. Nonetheless, integrating the O1, O2, and E2 terminations in \glspl{rsu} would be a relatively straightforward operation, as the communication between the devices and the \glspl{ric} in the \gls{ecc} could take place by using the already existing \gls{rsu} control plane. Similarly to the \gls{bs} case, proper extensions to the \gls{oran} interface definitions will enable \glspl{rsu} to be subjected to data collection and control. 

On the other hand, direct communication between the \glspl{ric} and the \glspl{cav} results in being more challenging. However, allowing the \gls{nr} Sidelink stack of \glspl{cav} to be centrally controlled opens the opportunity to address a vital issue in \gls{v2x} networks: orchestrating the \gls{v2v} communication. 

\subsection{\Gls{oran}-based control plane for \gls{v2v}}
In our integrated architecture proposition, we envision a \gls{vanet} retaining its decentralized nature while being supported by a next-generation \gls{oran} to enhance its performance to the point of supporting the stringent requirements of the safety-critical \gls{its} services. As Figure~\ref{fig:ric} shows in the exemplary scenario on the left, the data plane of the \gls{v2v} ad-hoc network is represented by direct \gls{nr} sidelink \gls{mmwave} connections established among \glspl{cav}, offering high throughput without occupying \gls{bs} resources. For the control plane, \gls{sub6} 5G links are employed, allowing for reliable and efficient communication between vehicles and the centralized controller through the existing \gls{ran} infrastructure. Using \gls{sub6} frequencies in the control plane offers a broader coverage and better penetration capabilities, leveraging on the ubiquitous coverage of modern cellular network deployments. The role of this out-of-band control plane is to relay \gls{oran} messages between the \glspl{ric} and the interface terminations of the \gls{cav}\footnote{In Figure~\ref{fig:ric}, only the E2 termination is included in \glspl{cav} for the sake of simplicity.}. According to this architecture, a \gls{cav} would require a \gls{5g} \gls{fr1} \gls{ue} to connect to the \gls{5g} \gls{bs} and an \gls{fr2} \gls{nr} Sidelink \gls{ue} to connect with other vehicles. After an \gls{fr1} connection is established with the \gls{bs}, dedicated radio bearers for the \gls{oran}-based management plane are established. In particular, at least one is required to create a \gls{gtp} tunnel with a \gls{upf} co-located with the \gls{ric} in the \gls{ecc}. As Figure~\ref{fig:ric} shows in detail, the \gls{gtp} tunnel can be then used to transparently connect the \gls{nrric} and the E2 termination in the \gls{cav}. Thanks to the capability of \gls{gtp} tunnels of maintaining IP endpoint connectivity through handovers, the high mobility of \gls{cav} is not expected to disrupt the proposed control plane. In other terms, the burden of managing the mobility of E2 terminations is left to the 5G connection, while \gls{oran} microservices obtain a reliable connection with the \glspl{cav} as they navigate through the coverage area. 

Alternatively to integrating E2 into \glspl{cav}, the sidelink signaling primitives such as the ones found in RRC, RLC, and MAC may be used to control the \gls{cav}'s communication \cite{tutorialBoban}. However, this approach is less flexible than allowing full control and telemetry through E2, O1 and O2. Indeed, Section~\ref{sec:opportunities} demonstrates that integrating \gls{oran} interfaces into vehicles, allowing full programmability of the vehicular communication stack, is crucial for effectively addressing the challenges of vehicular communication.
\subsection{Technological feasibility}
The proposed architecture is feasible from a technological realization standpoint, with minimal modifications. Integrating a \gls{5g} \gls{ue} and a Sidelink \gls{ue} into each vehicle allows for the establishment of both the \gls{fr1} connection with the \gls{5g} \gls{bs} and the \gls{fr2} \gls{nr} Sidelink connections with other vehicles. This integration can be achieved without significant challenges, as vehicles generally have flexible energy consumption constraints, making accommodating the necessary communication modules feasible.
Moreover, the proposed architecture does not require substantial modifications to the existing \gls{5g} stack, as it relies on standard \gls{5g} communication modes. 

However, although realizable, the architecture's effectiveness and performance must be thoroughly analyzed. Factors such as latency and control plane overhead (i.e., \gls{bs} resource utilization) must be carefully considered to ensure the architecture's viability. Such analysis should be conducted case-by-case, representing another open question in the context of \gls{oran} for \gls{v2x}. In the following section, we conduct a preliminary analysis of the effectiveness and viability of addressing key \gls{v2x} challenges through the proposed \gls{oran}-enabled architecture.

\section{An O-RAN approach to the V2X challenges} \label{sec:opportunities}
\begin{table}[b!] 
\footnotesize
\centering
\caption{{Simulation settings}} \label{table:simulation_parameters}
%\footnotesize
\begin{tabular}{ | c | c | p{42mm} |}
\hline 
\multicolumn{2}{|c|}{Simulation parameters} \\
\hline
FR2 Frequency & 28 GHz\\
Max EIRP & 23 dBm  \\
Antenna Type & Uniform Planar Array \\
Simulation time & 300 s \\
Urban Pathloss model & 3GPP and ITU \cite{tutorialBoban, LinsalataLoSmap}  \\
Vehicular traffic density &  50-70 veh/km \\
PHY \& MAC & Sidelink \cite{tutorialBoban}  \\  
\hline 
	\end{tabular}
\end{table}
In the following, we discuss some fundamental challenges of \gls{v2x}, highlighting the potential of \gls{oran}-based solutions to provide significant improvements. For every challenge, we implement such a solution as an xApp and simulate it in a realistic scenario, whose parameters are reported in Tab~\ref{table:simulation_parameters}.
Numerical results show that xApps can effectively address these challenges, maximizing the \gls{v2x} network performance.
\subsection{Beam selection and management}
In the context of \gls{v2x}, \gls{mmwave} provides the communication capabilities required to support most of the core concepts of \gls{its}~\cite{choi2016}.

To establish effective directional communication at \gls{mmwave}, beams have to be aligned both at the transmitter and the receiver during the initial access or in case of connectivity interruption due to a blockage event or high interference. These critical operations are costly in terms of beam training overhead and become even more challenging due to the relative mobility of the vehicles, thus requiring tight beam tracking~\cite{gu2022multimodality}. 

As such, traditional beam management mechanisms are considered inadequate, and \gls{v2x}-tailored solutions are required instead. Data-driven approaches, based on a fusion of reported vehicle positions and planned path, blockage prediction, urban layout information, and past successful beam alignment, are proven effective in providing fast beam alignment and tracking for vehicular communication. 
\begin{figure}[!t]
	\centering \includegraphics[width=0.98\textwidth]{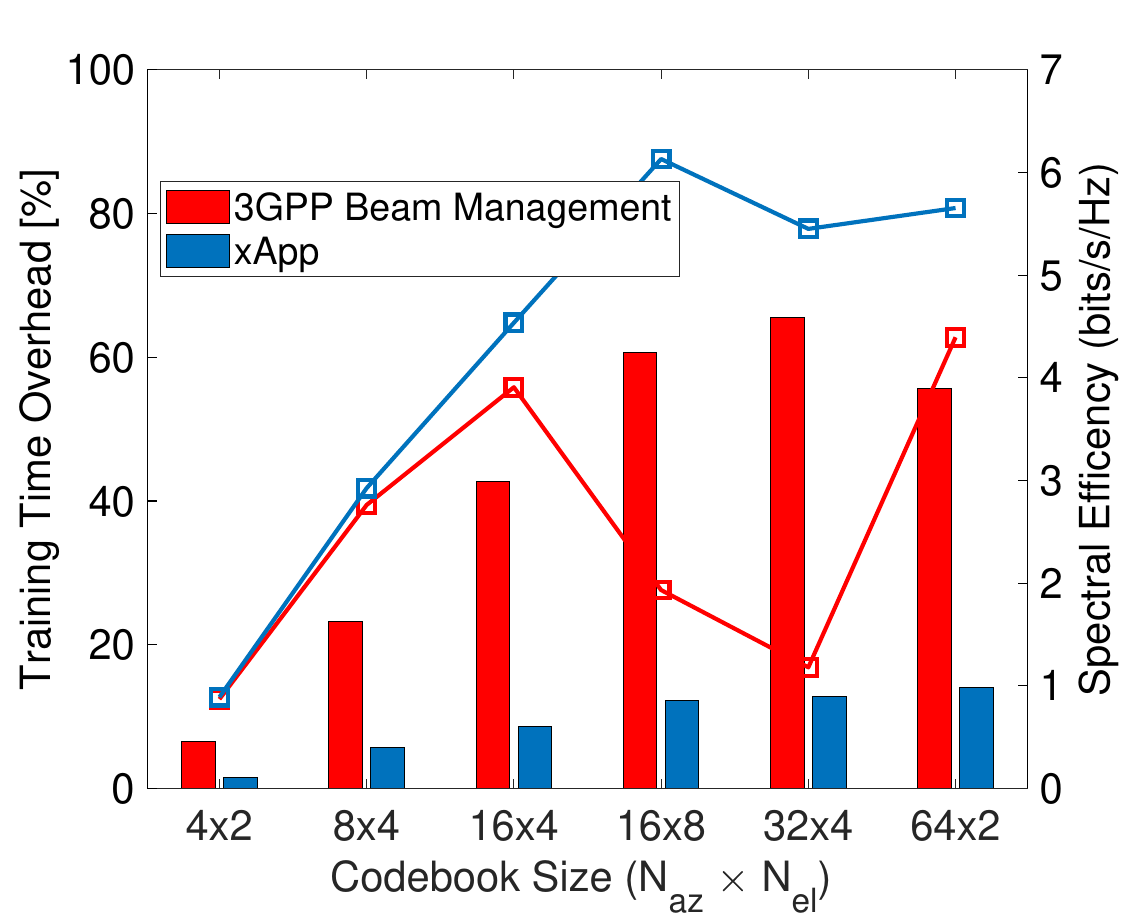}
	\caption{Beam Management training time overhead percentage (bars) and overall system spectral efficiency (lines) for different beam codebook cardinalities}
	\label{fig:BA_ov_se}
\end{figure}
While promising, such sophisticated solutions require access to a large amount of fresh network data from heterogeneous sources, which is hardly practical for the traditional \gls{ran} architecture. By contrast, thanks to the capability of abstracting the physical equipment idiosyncrasies and enabling large-scale data collection, \gls{oran} represents an ideal enabler for these solutions. 

Tapping into the wealth of up-to-date data exposed by the \gls{nrric}, an xApp could effectively host a well-informed beamforming management function based on arbitrary algorithms operating down to the millisecond timescale. Concurrently, an rApp running in the \gls{nrtric} can update beam management policies to fine-tune the overall objective of the beam management solution according to specific policies.

Overall, the potential of an \gls{oran}-based beam management has been proven in the most general settings~\cite{oranUsecases}. Nonetheless, there is still a lack of \gls{v2x}-dedicated studies on this matter where the capabilities of \gls{oran} are applied to beam management and massive-MIMO for vehicular communications. 

To support this claim, we implemented and simulated an xApp doing beam management using \gls{oran} control messages and timing, determining the subset of beams to test based on the latest \glspl{cav} positions and a priori environmental data. This method is compared with the gradient search used in \gls{3gpp}-standard 5G NR beam tracking \cite{tutorialBoban}, which searches around the optimal beam before any blockage or connectivity issues. Figure~\ref{fig:BA_ov_se} shows the results as a percentage of total training time relative to the connection duration and the average spectral efficiency. The results suggest that the proposed architecture can reduce system overhead (by up to 50\% for large codebook sizes) and improve spectral efficiency.

\subsection{Radio resource management.}
\begin{figure}[!t]
	\centering \includegraphics[width=0.98\textwidth]{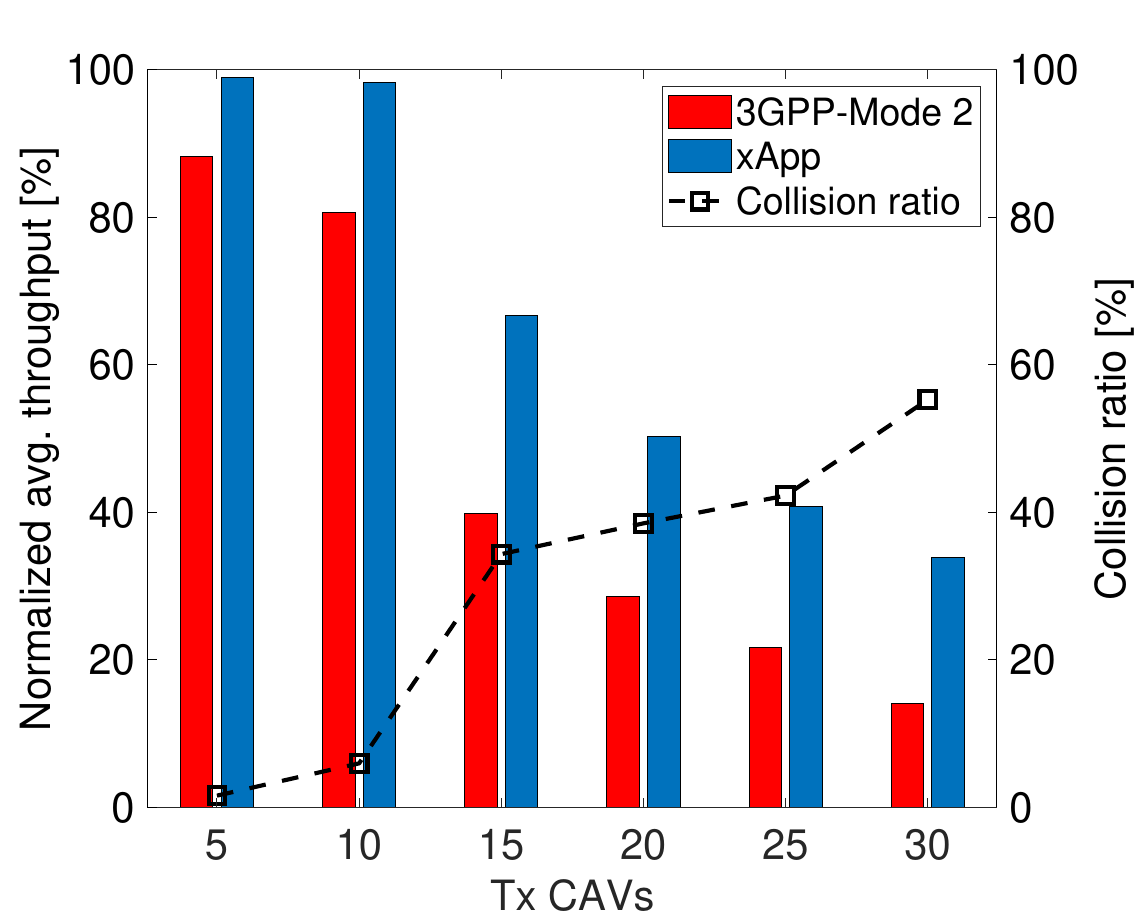}
	\caption{Average normalized throughput (bars) and the average collision ratio (line) versus the number of simultaneously transmitting \glspl{cav}}
	\label{fig:mode2_vs_oran}
\end{figure}
\Gls{its} are characterized by a large set of diverse services that present extremely challenging communication requirements.
Therefore, efficiently managing the scarce radio resources in the \gls{ran} to support these diverse requirements is crucial. \Gls{5g} foresees network slices, briefly described as bundles of virtualized resources dedicated to providing specific connectivity services to a subset of network users. Owing to the possibility of activating flexible and service-tailored slices, network slicing is considered a natural enabler of the diverse \gls{v2x} use cases~\cite{campolo2017}. 

However, physical resources still need to be efficiently allocated so that slices can support the communication requirements, and slice isolation needs to be guaranteed. Static resource partitioning is hardly viable due to the fast-changing state of the wireless network. Dynamic slice resource allocation based on continuous monitoring of the network parameters is to be preferred. This is especially true for \gls{v2x} due to the aforementioned challenging environment \cite{tutorialBoban, 6GV2Xchallenges}. 

Thanks to its extensive data collection and control capabilities, \gls{oran} is the fundamental enabler of slicing resource management~\cite{chih2020}. Nonetheless, the problem of practically enabling the slicing for \gls{v2x} in the real world has yet to be satisfactorily addressed. 

In this case, an \gls{oran}-based approach can fuse network measurements and external information (i.e., vehicle localization and planned path) to detect any potential criticality, reallocate slice resources accordingly, and allow for more efficient spectrum use~\cite{puligheddu2023semoran}. 

For instance, an xApp could monitor vehicular traffic and proactively allocate slice resources in those cells that will soon be subject to increased vehicular activity. If such resources are unavailable, xApps can unload the receiving cells by triggering handovers, reducing foreign slices' allotments, or disconnecting some users in the extreme. Additionally, by dynamically controlling \gls{sps} and \gls{cg}, the xApp can guarantee resource-efficient low latency and reliable communications both at the slice and single vehicle levels, ultimately providing isolation for safety-critical services. 

Given the complexity of the problem, \gls{ml}-based approaches are likely to be required. In this case, an rApp could monitor and fine-tune the mechanism put in place by the xApps and build or retrain appropriate \gls{ml} models, further adapting the \gls{v2x} slicing management to long-term environmental variations. 
Handover anomalies caused by complex vehicular mobility represent another challenge that can be effectively tackled by \gls{oran} apps, and this has been detailed as a particular use case by the \gls{oran} Alliance~\cite{oranUsecases}. 

The problem of resource allocation is also relevant for direct \gls{v2v} connections. 
According to the \gls{3gpp} standard, a central entity (i.e., a \gls{bs} or a \gls{rsu}) is expected to allocate the radio resources \cite{tutorialBoban}. This mechanism is hindered by the limited perception of the central entity with respect to each \gls{v2v} link condition and traffic requirement \cite{ShinRA2023}. Alternatively, a distributed resource allocation mechanism, namely Mode 2, is available in the standard~\cite{tutorialBoban}. Mode 2 does not suffer from the same limited-perception issue but generally shows limited performance due to its distributed nature. 

In this context, an xApp could gather data about the vehicle's position and mobility, channel status and interference profile. This information can be processed to adapt the allocation strategies to the fast-varying vehicular environment. 

Figure~\ref{fig:mode2_vs_oran} numerically proves the effectiveness of such an xApp by comparing it with Mode 2 resource allocation. In brief, Mode 2 consists of 2 phases: a first sensing phase, where the vehicle senses for free resources, and a second phase, where the vehicle transmits in a subset of the free-sensed resources. On the other hand, the implemented xApp receives bandwidth requests from the vehicles, schedules resources according to a proportional-fair mechanism, and transmits the proper grants back to the vehicles using the control plane of the proposed architecture.

Both solutions are simulated in the same scenario of Figure~\ref{fig:mode2_vs_oran}. The normalized average throughput is shown on the left Y-axis. The xApp approach generally guarantees higher throughput. This is especially true for the case of a large number of transmitting \glspl{cav}, where Mode 2 shows less than 50\% of the xApp throughput. This is mainly due to the distributed nature of Mode 2, which causes collisions (i.e., 2 or more vehicles utilizing the same slot). The collision ratio of Mode 2 plotted against the right Y-axis numerically confirms this. Note that no resource utilization collision happens in the xApp-based approach.  
\subsection{Enhanced Vehicle-to-vehicle connectivity.}

Fundamental \gls{its} services such as cooperative awareness, augmented reality, and coordinated autonomous driving require extensive data exchange among \glspl{cav} that are in close proximity to each other \cite{LinsalataLoSmap}. However, relying on base stations to forward the entirety of this traffic is impractical due to the resulting inefficiency and increased burden on the traditional \gls{ran} infrastructure. High-frequency Sidelink communications enabled by 5G \gls{nr} are thus inevitable for reliable low latency and high throughput \gls{v2v} links.

The high mobility of the vehicles and the harsh \gls{mmwave} propagation creates a challenging twist on the traditional problems of Ad-hoc networks, which include link selection, network graph and routing optimization, and congestion control. By tapping into the control and monitoring capabilities of the \gls{oran} architecture, an xApp could select which \gls{v2v} links to activate according to the expected channel quality and probability of blockage. 
The overall link selection strategy could optimize the underlying \gls{vanet} graph to meet different objectives. For instance, different superimposed graphs - low latency graphs prioritizing short paths or high throughput graphs prioritizing high-quality links - can be precomputed and dynamically activated according to the instantaneous communication needs or the policies dictated by an rApp. 

To address the problem defined above, we implemented an xApp that optimally selects relays (namely \glspl{rsu} or other \glspl{cav}) among those available to establish a multi-hop path between two vehicles whose direct link is in a \gls{nlos} condition. At first, the xApp reconstructs the vehicular network graph based on the quality of all the \gls{v2v} links that can be established at that moment. Such information can be obtained, for instance, by knowing the relative position of the vehicles and their communication characteristics or through periodic reports via the control plane. Then, the xApp applies a linear optimization method to select the relayed path that minimizes hops while guaranteeing a minimum end-to-end \gls{snr}. 
\begin{figure}[!t]
	\centering \includegraphics[width=0.98\textwidth]{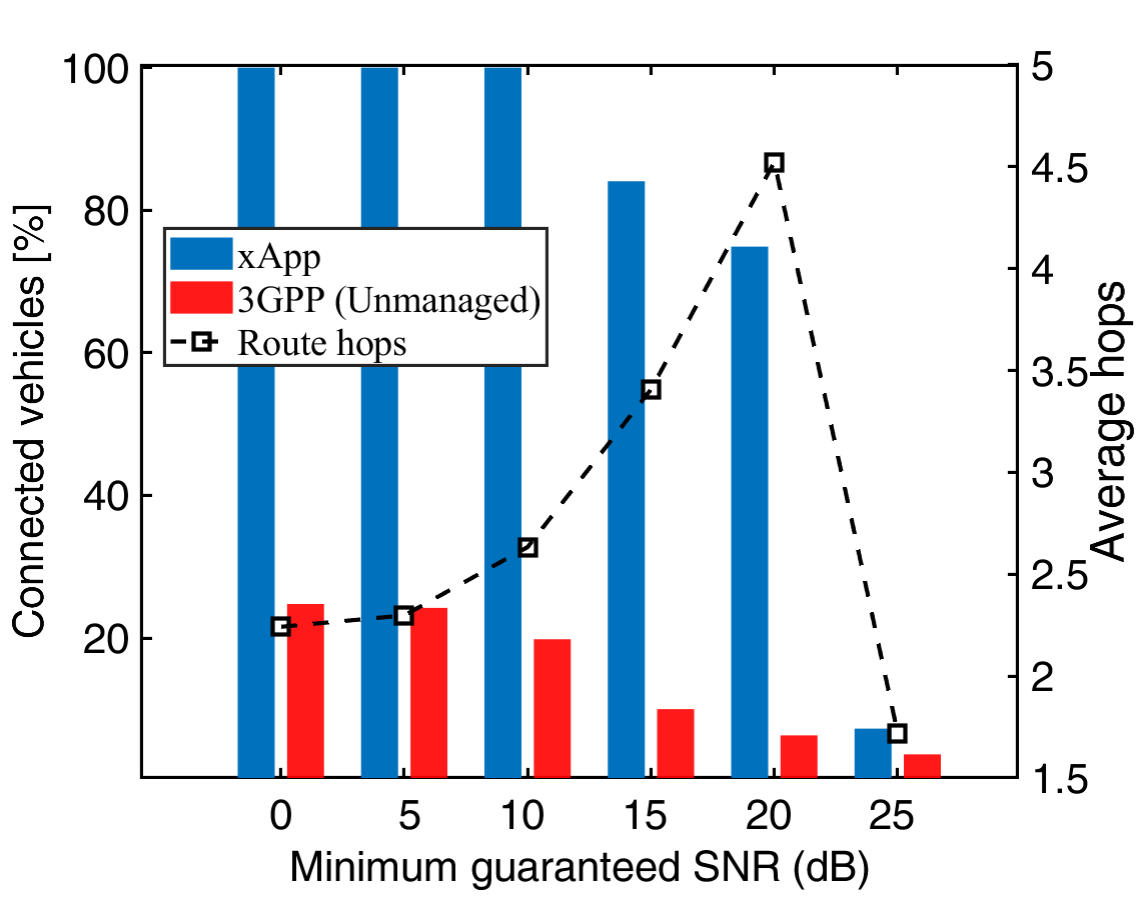}
	\caption{Network connectivity (bars) and average number of hops (line) versus required SNR}
	\label{fig:new_barplot}
\end{figure}
Results are shown in Figure~\ref{fig:new_barplot}. The baseline approach, which considers only the direct links, shows that no more than 25\% of the vehicles can establish a connection throughout the entire time window. This KPI decreases when the minimum \gls{snr} guarantees become more stringent. On the other hand, it is shown how the simulated xApp can guarantee full vehicular connectivity even for high levels of minimum guaranteed \gls{snr}. 
Figure~\ref{fig:new_barplot} also plots the average number of hops required to ensure vehicular connectivity. This measure shows a trade-off between minimum \gls{snr} - affecting the throughput - and path length - affecting the latency - which could be further exploited in an alternative xApp. 
\begin{figure}[!t]
	\centering \includegraphics[width=0.98\textwidth]{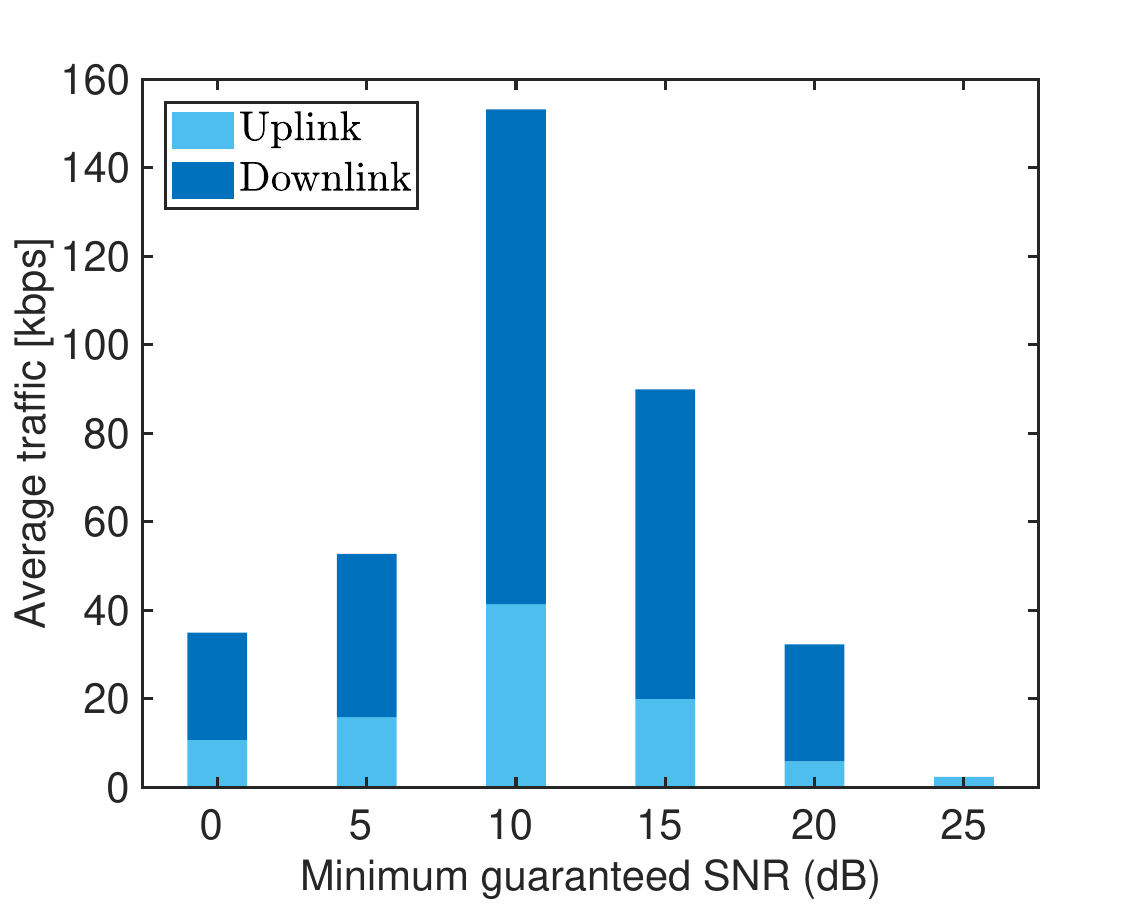}
	\caption{Control plane traffic required by the \gls{v2x} xApp}
	\label{fig:overhead}
\end{figure}

Among all the xApps proposed in this section, the relay selection xApp presents the highest control plane traffic. Consequently, an analysis of the control plane overhead is provided. For every blockage event, the proposed xApp requires two control messages in uplink - signaling a link failure and a relay path request - and as many downlink control messages as the number of hops. 
By considering every blockage event, figure~\ref{fig:overhead} reports the cumulative volume of control plane traffic averaged over the 5-minute time window and with a pessimistic \gls{oran} packet size of $1\,$Kb. As expected, the downlink traffic constitutes the most significant part of the control plane overhead. Nonetheless, the worst-case traffic of $160\,$Kbps is negligible for a 5G connection. This confirms the feasibility of the proposed xApp in terms of communication overhead. 

\begin{figure}[!t]
	\centering \includegraphics[width=0.98\textwidth]{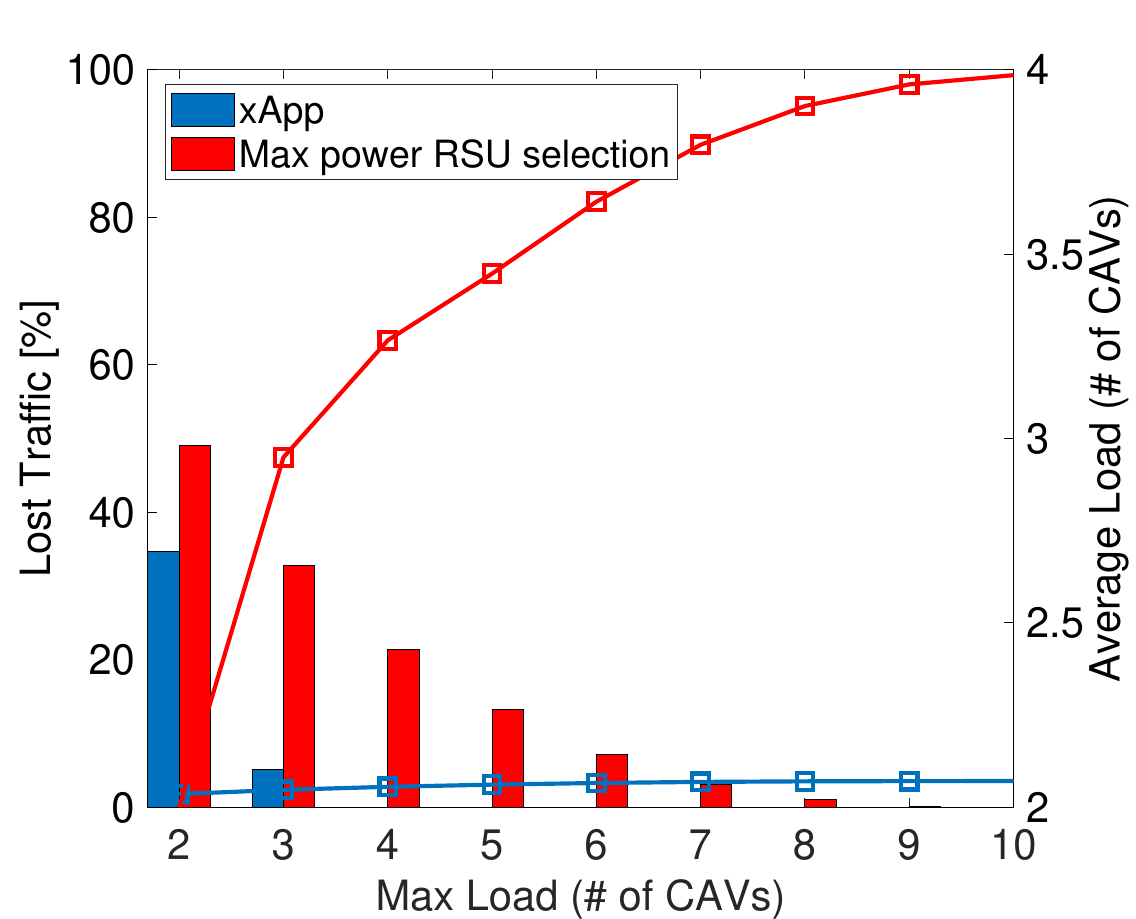}
	\caption{Lost traffic (bars) and average load (lines) versus RSU maximum number of simultaneously supported \glspl{cav} }
	\label{fig:rsu_selection}
\end{figure}
Finally, a reactive xApp like the one proposed must be able to apply the control solution fast enough to be effective.  
In the proposed system, the control latency boils down to the latency of the 5G system supporting the control plane, which we fix to a conservative $30\,$ms. 
The simulation statistics show how the blockage duration is centered around values going from $3$ to $10\,$ seconds. Therefore, the control plane latency is capable of supporting the scenario's dynamics, and ample margin is left to the xApp for the best path computation.
\subsection{Programmable and up-to-date V2X digital twin.}
To obtain an accurate real-time digital reproduction of the physical environment, a \gls{dt}-enabled \gls{v2x} system has to use high-definition 3D maps and combine multi-modal sensory data from several vehicles' onboard sensor data, as well as a detailed description of the communication network state.  
The \gls{oran} architecture is well-positioned to source the network information required to build high-fidelity \gls{v2x} \glspl{dt}, reducing the amount of data that the network nodes should manage and exchange. 
At the same time, \gls{oran} applications can exploit the \gls{dt} itself to run inference on the overall \gls{v2x} scenario without causing communication overhead with the infrastructure.  
rApps can retrain \gls{ml} models on the virtual \gls{v2x} environment recreated by the \gls{dt} to ensure that the xApp data-driven approaches always employ up-to-date models.

Proactive and optimized traffic forecasting can be leveraged from the \gls{v2x} \gls{dt}. Figure \ref{fig:rsu_selection} shows the percentage of lost traffic due to \glspl{rsu} simultaneously serving \glspl{cav} under different load capabilities constraints. The standard \glspl{rsu} selection scheme, which maximizes received power, is compared with an xApp that considers power conditions and \gls{rsu} traffic forecasting. The xApp outperforms by minimizing lost traffic and average \gls{rsu} load, thanks to its ability to optimize \gls{rsu} allocation and selection using the \gls{dt} information.

\section{Concluding Remarks}
\label{sec:conclusion}

The increasing need for reliable and efficient vehicular communication is paramount in today's connected world. This paper pioneered the integration of \gls{oran} with \gls{v2x} communications for \gls{5g} and future \gls{6g} networks, highlighting \gls{oran}'s potential as a flexible, scalable, and cost-effective solution.
An architecture that tightly integrates \gls{v2x} and \gls{oran} was proposed, addressing key \gls{v2x} challenges with \gls{oran}-based solutions. Simulations demonstrated the benefits of a managed, programmable \gls{v2x} system, showing superior performance of the implemented xApps in terms of throughput, network resilience, and reduced control overhead compared to standard procedures.
These findings validate the proposed architecture and confirm \gls{oran}'s potential to support \gls{v2x} communications, paving the way for future innovations in intelligent networks and automated transport systems.

\bibliographystyle{IEEEtran}
\bibliography{biblio.bib}

\newpage
\section*{Biographies}
\vskip -2\baselineskip plus -1fil
\begin{IEEEbiographynophoto}{Eugenio Moro} is a researcher at Politecnico di Milano. His research area is wireless networks, with a focus on optimization, programmability and smart propagation. 
\end{IEEEbiographynophoto}
\vskip -2\baselineskip plus -1fil
\begin{IEEEbiographynophoto}{Francesco Linsalata} is a researcher at Politecnico di Milano. His main research interests focus on V2X communications and the next generation of wireless networks. 
\end{IEEEbiographynophoto}
\vskip -2\baselineskip plus -1fil
\begin{IEEEbiographynophoto}{Maurizio Magarini} is an Associate Professor at Politecnico di Milano. His research interests are in the broad area of communication and information theory, with focus on molecular communications, massive MIMO, vehicular communications, study of advanced waveforms for 6G, and wireless networks using unmanned aerial vehicles and high-altitude platforms. 
\vskip -2\baselineskip plus -1fil
\begin{IEEEbiographynophoto}{Umberto Spagnolini} is Professor of Statistical Signal Processing, Director of Joint Lab Huawei-Politecnico di Milano and Huawei Industry Chair, scientific coordinator of 6G Wireless Networks and Technologies, a large Eu-National project. %He is author of more than 380 papers on peer-reviewed journals/conferences and patents. 
Recent interests are on MIMO channel estimation, cooperative and distributed inference, vehicular systems (V2X and radar), integrated communication and sensing. %He is technical experts of standard-essential patents and IP.
\end{IEEEbiographynophoto}
\end{IEEEbiographynophoto}
\vskip -2\baselineskip plus -1fil
\begin{IEEEbiographynophoto}{Antonio Capone} (Fellow, IEEE) is currently the Dean of the School of Industrial and Information Engineering, Politecnico di Milano (Technical University of Milan). His main research activities include radio resource management in wireless networks, traffic management in software defined networks, network planning, and optimization. 
\end{IEEEbiographynophoto}

\end{document}